\documentclass[twocolumn,showpacs,preprintnumbers,superscriptaddress,amsmath,amssymb]{revtex4}%
\usepackage{amsfonts}
\usepackage{amsmath}
\usepackage{amssymb}
\usepackage{graphicx}
\usepackage{footmisc}%
\begin{document}
\title{Dynamical interpretation of the wavefunction of the universe}
\author{Dongshan He}
\affiliation{State Key Laboratory of Magnetic Resonances and Atomic and Molecular Physics,
Wuhan Institute of Physics and Mathematics, Chinese Academy of Sciences, Wuhan
430071, China}
\author{Dongfeng Gao}
\affiliation{State Key Laboratory of Magnetic Resonances and Atomic and Molecular Physics,
Wuhan Institute of Physics and Mathematics, Chinese Academy of Sciences, Wuhan
430071, China}
\author{Qing-yu Cai}
\thanks{Corresponding author. Electronic address: qycai@wipm.ac.cn}
\affiliation{State Key Laboratory of Magnetic Resonances and Atomic and Molecular Physics,
Wuhan Institute of Physics and Mathematics, Chinese Academy of Sciences, Wuhan
430071, China}

\begin{abstract}

In this paper, we study the physical meaning of the wavefunction of the universe.
With the continuity equation derived from the Wheeler-DeWitt (WDW) equation in the minisuperspace model,
we show that the quantity $\rho(a)=|\psi(a)|^2$ for the universe is inversely proportional to the
Hubble parameter of the universe.
Thus, $\rho(a)$
represents the probability density of the universe staying in the state $a$ during its evolution,
which we call the dynamical interpretation of the wavefunction of the universe.
We demonstrate that the dynamical interpretation can predict the evolution laws of the universe
in the classical limit as those given by the Friedmann equation.
Furthermore, we show that the value of the operator ordering factor $p$ in the WDW equation
can be determined to be $p=-2$.

\end{abstract}

\maketitle

\section{Introduction}

In quantum mechanics, the state of a particle is completely described by
its wavefunction $\Psi(x,t)$. At the time $t$, the probability of finding the particle
in an interval $\Delta x$ about the position $x$ is proportional to
$|\Psi(x,t)|^{2}\Delta x$, and thus $\rho(x,t)=|\Psi(x,t)|^{2}$ is interpreted
as the probability density and $\Psi(x,t)$ is called the probability amplitude
for the particle.
This is the statistical interpretation or the ensemble interpretation of the
wavefunction determined by the Schr\"{o}dinger equation in the standard quantum mechanics.
In quantum cosmology theory, the universe is described by a wavefunction
$\psi(h_{ij},\phi)$ determined by
the quantum gravity equation, called the Wheeler-DeWitt (WDW) equation~\cite{bd67},
$\mathcal{H}\psi(h_{ij},\phi)=0$, where $h_{ij}$ is the 3d metric and $\phi$ is
a scalar field, rather than the classical spacetime.
In principle, the wavefunction should contain all information about the
universe~\cite{hh83}, although it is hard to extract all the information from
it~\cite{av94}.
At first glance, the wavefunction of the universe should satisfy the statistical
interpretation~\cite{ch91}.
But, if we adopt the statistical interpretation for the wavefunction of the universe directly,
it would be strange that the quantity $|\psi(h_{ij},\phi)|^2$ denotes
the probability density of finding a universe somewhere.

The observed universe is unique.
If we want to study
the physical meaning of the wavefunction of the universe with only one universe,
the statistical or ensemble interpretation should be abandoned. In this paper, we
study the properties of the wavefunction of the universe in the minisuperspace model,
in which the wavefunction of the universe can be determined by the unique parameter,
the scale factor $a$ to describe the time-dependent evolution of the universe.
First of all, we find that the operator ordering factor $p$ representing the ambiguity
in the ordering of noncommuting operators in the WDW equation should take value $p=-2$
due to the requirement of finiteness of the wavefunction of the universe.
Next, we show that the quantity $\rho(a)=|\psi(a)|^2$ for the universe is inversely
proportional to the Hubble parameter of the universe, and represents the probability density
of the universe staying in the state $a$ during its evolution.
We further show that the dynamical interpretation of the wavefunction of the universe can
give inflation solutions of the small universe and the correct evolution laws of the universe
in the classical limit, as required by the correspondence principle.
This paper is organized as follows. In Sec.~\ref{wdwe:mini:model}, the WDW equation is applied
to the minisuperspace model. Then, the formula of $\rho(a)$ for the universe is obtained with a
determined $p$.
in Sec.~\ref{formula:density:matrix}. The dynamical interpretation of the
wavefunction of the universe is given in Sec.~\ref{dynamical:interpretation:wavefunction:universe}.
The correct evolution laws of the universe in the classical limit with dynamical
interpretation is discussed in Sec.~\ref{evolution:universe:classical:limit}.
In Sec.~\ref{dynamical:interpretation:scalar:field}, the formula of $\rho(a)$ of the
universe within the scalar field model is obtained. Finally, we discuss and conclude
in Sec.~\ref{discussion:conclusion}.

\section{WDW equation in the minisuperspace model}
\label{wdwe:mini:model}

Assumed to be homogeneous and isotropic, the universe can be
described by a minisuperspace model~\cite{npn00,npn12,apk97} with one single
parameter, the scale factor $a$.
The Einstein-Hilbert action for the model can be written as
\begin{equation}
S=\int\left(  \frac{Rc^{3}}{16\pi G}-\frac{9\varepsilon_{n}}{16c}\right)
\sqrt{-g}d^{4}x, \label{action}
\end{equation}
where the $\varepsilon_{n}$ represents the energy density of the universe, and
the constant before $\varepsilon_{n}$ is chosen for later convenience. Since
the universe is homogeneous and isotropic, the metric of the universe in the
minisuperspace model is given by
\begin{equation}
ds^{2}=\sigma^{2}\left[  -N^{2}(t)c^{2}dt^{2}+a^{2}(t)d\Omega_{3}^{2}\right]
. \label{metric}%
\end{equation}
Here, $d\Omega_{3}^{2}=dr^{2}/(1-kr^{2})+r^{2}(d\theta^{2}+\sin^{2}\theta
d\phi^{2})$ is the metric on a unit three-sphere, $N(t)$ is an arbitrary lapse
function, and $\sigma^{2}=2/3\pi$ is a normalizing factor chosen for later
convenience. Note that $r$ is dimensionless and the scale
factor $a(t)$ has length dimension~\cite{Weinberg}. From Eq.~(\ref{metric}),
one can get the scalar curvature
\begin{equation}
\mathcal{R}=6\frac{\ddot{a}}{\sigma^{2}c^{2}N^{2}a}+6\frac{\dot{a}^{2}}{\sigma^{2}
c^{2}N^{2}a^{2}}+\frac{6k}{\sigma^{2}a^{2}}, \label{Ricci}%
\end{equation}
where the dot denotes the derivative with respect to the time, $t$.
Inserting Eqs.~(\ref{metric}) and (\ref{Ricci}) into Eq.~(\ref{action}), we can
get
\begin{align*}
S  &  =\int\frac{6\sigma^{2}Nc^{4}}{16\pi G}\left(  \frac{a^{2}\ddot{a}}%
{N^{2}c^{2}}+\frac{a\dot{a}^{2}}{N^{2}c^{2}}+ka-\frac{G\varepsilon_{n}a^{3}%
}{c^{4}}\right)  d^{4}x,\\
&  =\frac{6\sigma^{2}Nc^{4}V}{16\pi G}\int\left(  \frac{a^{2}\ddot{a}}%
{N^{2}c^{2}}+\frac{a\dot{a}^{2}}{N^{2}c^{2}}+ka-\frac{G\varepsilon_{n}a^{3}%
}{c^{4}}\right)  dt,\\
&  =\frac{Nc^{4}}{2G}\int\left(  -\frac{a\dot{a}^{2}}{N^{2}c^{2}}%
+ka-\frac{G\varepsilon_{n}a^{3}}{c^{4}}\right)  dt.
\end{align*}
The Lagrangian of the bubble can thus be written as
\begin{equation}
\mathcal{L}=\frac{Nc^{4}}{2G}\left(  ka-\frac{a\dot{a}^{2}}{N^{2}c^{2}}%
-\frac{G\varepsilon_{n}a^{3}}{c^{4}}\right) , \label{L}%
\end{equation}
and the
momentum $p_{a}$ can be obtained as
\[
p_{a}=\frac{\partial\mathcal{L}}{\partial\dot{a}}=-\frac{c^{2}a\dot{a}}{NG}.
\]
Taking $N=1$, the Hamiltonian is found to be
\begin{eqnarray}
\mathcal{H}&=&p_{a}\dot{a}-\mathcal{L} \notag\\
&=&-\frac{1}{2}\left(  \frac{Gp_{a}^{2}}{c^{2}a}+\frac{c^{4}ka}
{G}-\varepsilon_{n}a^{3}\right)  .\nonumber
\end{eqnarray}
In quantum cosmology theory, the evolution of the universe is completely
determined by its quantum state that should satisfy the WDW equation. With
$\mathcal{H}\Psi=0$ and $p_{a}^{2}=-a^{-p}\frac{\partial}{\partial
a}(a^{p}\frac{\partial}{\partial a})$, we get the WDW equation \cite{bd67,swh84,av94},
\begin{equation}
\left(\frac{1}{a^{p}}\frac{\partial}{\partial a}
a^{p}\frac{\partial}{\partial a}-ka^{2}+\varepsilon_{n}a^{4}\right) \psi(a)=0. \label{WDWE}
\end{equation}
Here, $k=1,0,-1$ are for spatially closed, flat and open universe,
respectively. The factor $p$ represents the uncertainty in the choice of
operator ordering.
For simplicity, we take $\hbar=G=c=1$ in this paper.

\section{The square modulus of the wavefunction for the universe}
\label{formula:density:matrix}

In order to study the physical meaning of the wavefunction of the universe,
we should discuss the square modulus of the wavefunction for the universe first.
Since there is only one parameter $a$ in
Eq.~(\ref{WDWE}),  the complex function $\psi(a)$
can be formally rewritten as
\begin{equation}
\psi(a)=R(a)\exp(iS(a)),\label{psi}
\end{equation}
where $R$ and $S$ are real functions. Then the square modulus of the wavefunction is
\begin{equation}
\rho(a)\equiv\left\vert \psi(a)\right\vert ^{2}=R(a)^{2}. \notag
\end{equation}
It should be pointed out that, no matter what the physical meaning it is,
the value of $\rho(a)$ should be finite.

From Eq.~(\ref{WDWE}), it is
easy to construct a conserved current $j^a$ as~\cite{our14,av88}
\begin{eqnarray}
&&  j^{a}=\frac{i }{2}a^{p}(\psi^{\ast}\partial_{a}\psi-\psi\partial_{a}%
\psi^{\ast}),\label{ja1}\\
&&  \partial_{a}j^{a}=0.\label{ja2}%
\end{eqnarray}
Inserting Eq.~(\ref{psi}) into Eq.~(\ref{ja1}), we can get
\begin{equation}
j^{a}=-a^{p}R^{2}S^{\prime},\nonumber
\end{equation}
where the prime denotes the derivative with respect to $a$.
Integrating Eq.~(\ref{ja2}) gives that
\begin{equation}
a^{p}R^{2}S^{\prime}=c_0,\label{ja3}
\end{equation}
where the $c_0$ is a dimensionless integral constant.

In the quantum Hamilton-Jacobi theory, the relation between the action and the
momentum can be written as~\cite{lpg93,npn00},
\begin{equation}
p_{a}=\frac{\partial S}{\partial a}=\frac{\partial\mathcal{L}}{\partial\dot{a}}
=- a\dot{a}. \label{gr2}
\end{equation}
According to Eqs.~(\ref{ja3}) and (\ref{gr2}), we can get $\rho(a)$ for the universe
\begin{equation}
\rho(a)=R(a)^2=-\frac{c_0 }{a^{p+1}\dot{a}}. \label{rou}
\end{equation}
Using the definition of the Hubble parameter
\begin{equation}
H(a)=\frac{\dot{a}}{a},
\end{equation}
we can rewrite $\rho(a)$ as
\begin{equation}\label{dynamical:equation}
\rho(a)=-\frac{c_0 }{a^{p+2}H(a)}.
\end{equation}
In principle, the value of $\rho(a)$ should be finite for any $a$, i.e.,
no matter whether the universe is very small or large enough.
In both the inflation stage and the dark energy stage, the value of the Hubble parameter
should be finite. From
\begin{align*}
\rho(a\rightarrow 0)=-\frac{c_0 }{a^{p+2}H(a) },
\end{align*}
we can get a boundary condition
\begin{align*}
p+2\leq0.
\end{align*}
On the other hand, the finiteness of $\rho(a)$ with large $a$
\begin{align*}
\rho(a\rightarrow\infty)=-\frac{c_0 }{a^{p+2}H(a)}
\end{align*}
requires that
\begin{align*}
p+2\geq0.
\end{align*}
The above two boundary conditions determine the value of the ordering factor to be  $p=-2$.
Thus, Eq.~(\ref{dynamical:equation}) is reduced to
\begin{equation}\label{dynamical:interpretation}
\rho(a)=-\frac{c_0}{H(a)}.
\end{equation}
With the above formula at hand, we can discuss now the physical
meaning of the wavefunction of the universe.

\section{Dynamical interpretation of the wavefunction of the universe}
\label{dynamical:interpretation:wavefunction:universe}

In quantum mechanics, the wavefunction $\Psi(x,t)$ of a particle is
interpreted as probability amplitude and
$|\Psi(x,t)|^2\Delta x$ is the probability of finding the particle at time $t$ in the
interval $\Delta x$. This is the statistical interpretation or the ensemble
interpretation of wavefunction. However, such an interpretation of wavefunction
in quantum mechanics cannot be applied to the wavefunction of the universe in
quantum cosmology.
For an observer inside the universe, it is should be very strange if the wavefunction
is related to the probability of finding a universe. So, the physical meaning of
the wavefunction for the universe should be reinterpreted.

From Eq.~(\ref{dynamical:interpretation}), we can see that the value of
$\rho(a)$ for the universe only depends on the Hubble parameter $H(a)$. $\rho(a)$ is 
large when the universe expands slowly,
and it is small when the universe expands quickly.
So, it is obvious that $\rho(a)$ for the universe represents the
evolution velocity of the universe and thus relates to the dynamics of the universe.
In this case, $\rho(a)$ can be treated as the dynamical interpretation.
Similar to the statistical interpretation of the wavefunction in quantum mechanics,
the dynamical interpretation for the wavefunction of the universe can be explicitly
described as:
\begin{center}
$\int_{a_1}^{a_2}\psi^*(a)\psi(a)da$ is proportional to the time spent when the universe
involves from the state $a_1$ to $a_2$.
\end{center}
In this way, we have showed that the physical meaning of $\rho(a)$ for the
universe can be interpreted as the probability density of the universe staying in
the state $a$ during its evolution.

Generally speaking, an interpretation for the physical meaning of wavefunction should
satisfy the correspondence principle, i.e., the quantum cosmology can reduce to the
classical cosmology in the classical limit within the dynamical interpretation.
In fact, the dynamical interpretation completely depends on the evolution equation Eq.~(\ref{rou}),
which we call the dynamical equation for the universe.
As will be shown below, solutions of the WDW equation in the classical limit ($a\gg 1$)
together with the dynamical equation can predict the same evolution laws of the universe
as those given by the Friedmann equation.
The exponential expansion solutions of the early universe ($a\ll 1$) can also be obtained
from the WDW equation together with the dynamical equation~(\ref{rou}).

\section{The evolution of the universe with the dynamical equation}
\label{evolution:universe:classical:limit}

If we want to probe the rationality of
the dynamical interpretation of the wavefunction of the universe, we should
verify whether the dynamical interpretation can give the correct evolution laws
of the universe in the classical limit or not~\cite{av89}. On the other hand,
the quantum behaviors of the small enough universe should be reserved with
the dynamical interpretation.
The classical evolution laws of the universe at different stages dominated
by radiation, matter and dark matter, respectively, can be obtained by
solving the Friedmann equation. The evolution equations of the classical universe
at different stages are shown in Tab.~\ref{table1}.
\begin{table}[!h]
\tabcolsep 5mm
\begin{center}
\begin{tabular}{l@{}ll@{}ll@{}l}
\hline
\multicolumn{2}{l}{dominator}&\multicolumn{2}{l}{density}&\multicolumn{2}{l}{evolution} \\ \hline
radiation&   &n=4&  &$a(t)\propto(t+t_{0})^{1/2}$& \\
matter&      &n=3&   &$a(t)\propto(t+t_{0})^{2/3}$& \\
dark energy&  &n=0&  &$a(t)\propto e^{t}$& \\
\hline
\end{tabular}
\end{center}
\caption{The classical evolution laws of the universe given by the Friedmann equation.
The universe is dominated by different kinds of energies $\varepsilon_n=\lambda_n/a^n$
at different stages.}
\label{table1}
\end{table}

Let us study the evolution laws of the universe with the dynamical interpretation
of the universe. For simplicity, we only consider the case of the flat
universe $k=0$.
The WDW equation for the flat universe with energy density
$\varepsilon_{n}$ can be written as
\begin{equation}
\left(  \frac{1}{a^{p}}\frac{\partial}{\partial a}a^{p}\frac{\partial
}{\partial a}+\varepsilon_{n}a^{4}\right)  \psi(a)=0.\label{WDWEn}
\end{equation}
Here $\varepsilon_{n}=\lambda_{n}/a^{n}$, for $n=4,3,0$ representing the
universe dominated by radiation, matter and dark energy,
respectively~\cite{dhc05}.

In principle, the universe contains all kinds of energies at the same
time, so the energy density should take the form of $\varepsilon=\varepsilon
_{0}+\varepsilon_{3}+\varepsilon_{4}$.
In practice, the universe is dominated by one kind of energy $\varepsilon_{n}$
at a specific stage. During the evolution of the universe, $n$ changes slowly
from $n=4$ to $n=0$. For an arbitrary $n$,
we can obtain the analytical solutions of Eq.~(\ref{WDWEn}),
\begin{eqnarray}
\psi_{n}(a) =a^{\frac{1-p}{2}} \left[ {i c_{1}J_{\nu}\left(
\frac{\sqrt{\lambda_{n}}a^{3-n/2}}{3-n/2}\right)}\right. \notag \\ \left.{ +c_{2}Y_{\nu}
\left( \frac{\sqrt{\lambda_{n}}a^{3-n/2}}{3-n/2}\right)}\right].
\label{wave:function}
\end{eqnarray}
Here, $J_{\alpha}(x)'s$ are Bessel functions of the first kind, $Y_{\alpha}(x)'s$
are Bessel function of the second kind, and $\nu=\left\vert(1-p)/(n-6)\right\vert$.

First, consider the wavefunction of the WDW equation in the classical limit ($a\gg1$).
For $x\gg\left\vert
\nu^{2}-1/4\right\vert $, Bessel functions take the following asymptotic
forms,
\begin{align*}
J_{\nu}(x)  &  \sim\sqrt{\frac{2}{\pi x}}\cos(x-\nu\pi/2-\pi/4),\\
Y_{\nu}(x)  &  \sim\sqrt{\frac{2}{\pi x}}\sin(x-\nu\pi/2-\pi/4).
\end{align*}
If the free parameters $c_1$ and $c_2$ in Eq.~(\ref{wave:function})
take the values of $c_{1}=c_{2}=c_{-}\sqrt{\pi/2}$, the wavefunction can
be rewritten as
\begin{equation}
\psi_{n}(a)=c_{-}a^{\frac{n-2p-4}{4}}\exp\left[  \frac{-i\sqrt{\lambda_{n}%
}a^{3-n/2}}{3-n/2}+i\theta\right], \label{psaiV}
\end{equation}
where $\theta=(3n-2p-16)\pi/(4n-24)$, and  $S^{\prime}<0$.
From Eq.~(\ref{gr2}), we know that the wavefunction in Eq.~(\ref{psaiV})
describes an expanding universe as suggested by Vilenkin~\cite{av88}.
With the wavefunction above, we can get
\begin{align*}
\rho(a)\equiv\left\vert \psi(a)\right\vert ^{2}=c_{-}^{2}a^{-p-2+n/2}.
\end{align*}
Together with the dynamical equation (\ref{rou}), the above equation gives
\begin{equation}\label{scale:factor}
\dot{a}=\frac{-c_0}{c_{-}^{2}a^{-1+n/2}},
\end{equation}
where $c_0<0$. Transforming the formula in Eq.~(\ref{scale:factor}) into
\begin{equation}
a^{-1+n/2}da=\frac{-c_0}{c_{-}^{2}}dt, \notag
\end{equation}
and then integrating both sides of the equation, we have
\begin{equation}
a(t)\propto\left\{  \begin{aligned}
&(t+t_{0})^{2/n},\,\,\,\,\,\,\,\,\,\,\,\,\,\,\,\,\,\,\,\, n\neq 0, \\
&e^{t+t_0},\,\,\,\,\,\,\,\,\,\,\,\,\,\,\,\,\,\,\,\,
\,\,\,\,\,\,\,\,\,\,\,\,\,\,\,  n=0. \\
\end{aligned}\right.
\notag
\end{equation}
The evolution laws of the universe from the WDW equation in the classical limit
($a\gg 1$) are completely consistent with the solutions of the Friedmann equation
as shown in Tab.~\ref{table1}.
It is interesting that the evolution equation of the universe derived from the
WDW equation is independent of the operator ordering factor $p$, which definitely
means that $p$ only represents the quantum effects of the universe~\cite{our14}.

Next, consider the evolution of the early universe ($a\ll 1$) within the dynamical
interpretation.
For $x\ll\left\vert \nu^{2}-1/4\right\vert $, Bessel functions take the
following asymptotic forms,
\begin{align*}
J_{\nu}(x)  &  \sim\left(  \frac{x}{2}\right)  ^{\nu}\frac{1}{\Gamma(\nu
+1)},\\
Y_{\nu}(x)  &  \sim-\frac{\Gamma(\nu)}{\pi}\left(  \frac{x}{2}\right)  ^{-\nu
}.
\end{align*}
So the wavefunction Eq.~(\ref{wave:function}) for the small scale factor ($a\ll 1$)
can be rewritten as
\begin{align*}
\psi_{n}(a)=c_{-}\sqrt{\frac{\pi}{2}}\left(  \frac{i\lambda_{n}^{\nu/2}%
a^{1-p}}{\Gamma(\nu+1)(6-n)^{\nu}}-\frac{\Gamma(\nu)(6-n)^{\nu}}{\pi
\lambda_{n}^{\nu/2}}\right).
\end{align*}
Here, we have assumed that $p<1$, otherwise the wavefunction is divergent.
$c_{1}$ and $c_{2}$ take the same value $c_{-}\sqrt{\pi/2}$ as that determined
by the solutions in the classical limit. In this case, the probability density of
the early universe is
\begin{align*}
\rho(a\ll 1)\equiv\left\vert \psi_{n}(a\ll 1)\right\vert
^{2}=c_{-}^{2}\frac{\Gamma^{2}(\nu)(6-n)^{2\nu}}{2\pi\lambda_{n}^{\nu}}.
\end{align*}
It is obvious that, when $a\ll 1$, $\rho(a)$ approximates a constant denoted as
$\rho_{0}(n)$.
When $a\ll 1$, the dynamical equation (\ref{rou}) can
be rewritten as
\begin{align*}
a^{p+1}\dot{a}  &  =-\frac{c_0}{\rho_{0}(n)},
\end{align*}
which is solved by
\[
a(t)=\left\{  \begin{aligned}
&\left[  -(p+2)c_0(t+t_{0})/\rho_{0}(n)\right]  ^{\frac{1}%
{p+2}},\,\,\,\,\,\,\,\,\,  p\neq-2, \\
&e^{H(t+t_{0})},\,\,\,\,\,\,\,\,\,\,\,\,\,\,\,\,\,\,\,\,\,\,\,\,\,\,\,
\,\,\,\,\,\,\,\,\,\,\,\,\,\,\,\,\,\,\,\,\,\,\,\,\,\,\,\,\,\,\,\,\,\,\,\,\,\,\,\,  p=-2, \\
\end{aligned}\right.
\]
where $H=-c_0/\rho_{0}(n)$.

When the universe is very small, its behaviors are dominated by quantum effects
and the evolution of the universe depends on the operator ordering factor $p$:
different $p$ gives different evolution equations of the scale factor $a(t)$.
In fact, the ordering factor
$p$ has been determined by the boundary conditions of finite density matrix of the universe
as $p=-2$.
This specific value $p=-2$ gives exponential expansion solutions
for the early universe, which is consistent with the result from quantum trajectory
theory~\cite{our14,our14a}.
Thus, we conclude that the WDW equation together with the dynamical interpretation can give
exponential expansion solutions of the early universe and the correct evolution laws of the
universe in the classical limit.

\section{The dynamical interpretation for the minisuperspace model with a scalar field}
\label{dynamical:interpretation:scalar:field}

In quantum cosmology, the most widely used model is the minisuperspace model
with a scalar field.
For a FRW universe filled with a scalar field $\phi$, the WDW equation
can be written as~\cite{av86,dlw00}
\[
\left[  \frac{1}{a^{p}}\frac{\partial}{\partial a}a^{p}\frac{\partial
}{\partial a}-\frac{1}{a^{2}}\frac{\partial^{2}}{\partial\phi^{2}}%
-U(a,\phi)\right]  \psi(a,\phi)=0,
\]
where $U(a,\phi)=a^{2}(k-a^{2}V(\phi))$.
We can rewrite the wavefunction $\psi(a,\phi)$ as $\psi(a,\phi)=R(a,\phi)e^{iS(a,\phi)}$,
where both $R(a,\phi)$ and $S(a,\phi)$ are real functions. Since the wavefunction $\psi(a,\phi)$ is
a function of $a$ and $\phi$, the currents for WDW equation can be obtained as~\cite{av86}
\begin{align}
j^{a}  &  =\frac{i}{2}a^{p}(\psi^{\ast}\partial_{a}\psi-\psi\partial_{a}%
\psi^{\ast}),\nonumber\\
&  =-a^{p}R^{2}(a,\phi)\partial_{a}S(a,\phi),\label{jaa}\\
j^{\phi}  &  =\frac{-i}{2}a^{p-2}(\psi^{\ast}\partial_{\phi}\psi-\psi
\partial_{\phi}\psi^{\ast}),\nonumber\\
&  =a^{p-2}R^{2}(a,\phi)\partial_{\phi}S(a,\phi). \label{jphi}%
\end{align}
The quantum Hamilton-Jacobi theory gives the guidance relations~\cite{npn12}
\begin{align*}
\partial_{a}S(a,\phi)  &  =-a\dot{a},\\
\partial_{\phi}S(a,\phi)  &  =a^{3}\dot{\phi}.
\end{align*}
The currents $j^{a}$ and $j^{\phi}$ satisfy the continuity equation,
\begin{equation}
\partial_{a}j^{a}+\partial_{\phi}j^{\phi}=0. \label{continuity}%
\end{equation}
Inserting Eqs.~(\ref{jaa}) and (\ref{jphi}) into Eq.~(\ref{continuity}), we can get
\[
\partial_{a}(a^{p}R^{2}\partial_{a}S)-\partial_{\phi}(a^{p-2}R^{2}
\partial_{\phi}S)=0.
\]
Applying the guidance relation to the equation above, we can obtain
\begin{equation}
\partial_{a}(a^{p}R^{2}\partial_{a}S)+\partial_{a}(a^{p-2}R^{2}\partial_{\phi
}S)a^{2}\dot{\phi}/\dot{a}=0. \label{continuity1}
\end{equation}
Integrating Eq.~(\ref{continuity1}) over $a$, we get
\[
a^{p+1}R^{2}\dot{a}-a^{p+3}R^{2}\dot{\phi}^{2}/\dot{a}+\int a^{p+1}R^{2}%
\dot{\phi}d(a^{2}\dot{\phi}/\dot{a})=-c_0.
\]
Let $A(a,\phi)=R^{-2}\int R^{2}a^{p+2}\dot{\phi}(2\phi_{a}+a\phi_{aa})da$, which gives
\begin{align}
\rho(a,\phi)=R^{2}(a,\phi)
=\frac{-c_0}{a^{p+1}\dot{a}(1-a^{2}\phi_{a}^{2})+A(a,\phi)},
\label{rouphi}
\end{align}
where$\ \phi_{a}=d\phi/da=\dot{\phi}/\dot{a}$, and $\phi_{aa}=d\phi_a/da$.
We can see when $\dot{\phi}\rightarrow0$, i.e., $A(a,\phi)\rightarrow0$
and $a^2\phi_{a}^2\rightarrow0$,
$\rho(\phi,a)$ in Eq.~(\ref{rouphi}) will
return to $\rho(a)$ in Eq.~(\ref{rou}), which indicates
that the dynamical interpretation still holds for the wavefunction of the early universe
with a slowly-rolling scalar field.

\section{Discussion and Conclusion}
\label{discussion:conclusion}

In summary, we have found the mathematical relation between the quantity $\rho(a)$
for the universe and the Hubble parameter that $\rho(a)$ is inversely proportional to
the Hubble parameter $H(a)$. We argue that $\rho(a)$ is not the probability density
of finding a universe somewhere, but represents the probability density of the universe staying
in the state $a$. This presents a dynamical interpretation of the wavefunction of the universe.
We have demonstrated that the dynamical interpretation can give the same evolution laws
of the universe as those given by the Friedmann equation, which satisfies the requirement of the
correspondence principle that the quantum cosmology should reduce to the classical cosmology
in the classical limit.
In this way, we have presented an investigation of the physical meaning of the
wavefunction of the universe.

Another result is that the value of the operator ordering factor $p$ that represents
the ambiguity in the ordering of noncommuting operators has also been determined.
With the requirement of the finiteness of the wavefunction of the universe,
the ordering factor should take value of $p=-2$.
This specific value of $p$ can predict exponential expansion solutions
of the small universe.
In fact, when the universe becomes large enough, the evolution of the universe is independent of the
value of the ordering factor $p$, which implies that $p$ represents the rules of the quantization
of the early universe and only dominates quantum behaviors of the universe.

\section*{Acknowledgement}

Finical support from the National Key Basic Research Program of China (NKBRPC) 
under Grant No. 2013CB922003 and National Natural Science Foundation of China 
(NSFC) under Grant No. 61471356 is gratefully acknowledged.

\end{document}